# Certificate Linking and Caching for Logical Trust

Qiang Cao    Vamsi Thummala    Jeffrey S. Chase    Yuanjun Yao    Bing Xie
Duke University
{qiangcao, vamsi, chase, yjyao, bingxie}@cs.duke.edu

**Abstract**

SAFE is a data-centric platform for building multi-domain networked systems, i.e., systems whose participants are controlled by different principals. Participants make trust decisions by issuing local queries over logic content exchanged in certificates. The contribution of SAFE is to address a key barrier to practical use of logical trust: the problem of identifying, gathering, and assembling the certificates that are relevant to each trust decision.

SAFE uses a simple linking abstraction to organize and share certificates according to scripted primitives that implement the application's trust kernel and isolate it from logic concerns. We show that trust scripting with logical data exchange yields compact trust cores for example applications: federated naming, nested groups and roles, secure IP prefix delegation and routing, attestation-based access control, and a federated infrastructure-as-a-service system. Linking allows granular control over dynamic logic content based on dependency relationships, enabling a logic server to make secure inferences at high throughput.

## 1. Introduction

Decentralized trust management (DTM [10]) has a long history (summarized in §2 and §7). In DTM systems, principals use a formal language to make statements about one another and about objects in the system, as the basis for access control and other trust decisions. DTM systems promise to enable rapid construction of a wide range of multi-domain networked systems on a common underlying foundation that is rigorous, flexible, and verifiable.

Researchers have produced many policy languages and authorization logics based on DTM principles (§7, §8), but these systems have seen little real-world practical use due to various practical challenges. We seek to develop a trust platform that can make future multi-domain DTM systems easier to develop and less brittle than today's systems. The key challenges are how to integrate the trust language with applications and how to identify, gather, and update relevant content. While second-generation DTM systems moved away from procedural policy languages in favor of declarative (logic) languages, applications have important procedural needs that are specific to how they use the logic.

Our premise is that these needs can be met by a programmable control layer above a simple trust data plane based on a standard logic language (Datalog). The control layer adds scripting primitives to organize logic content and direct the flow of logic content and queries. It produces and consumes logic content as certificates, and organizes these certificates according to the needs of the application.

Our system, SAFE, introduces *trust scripts* at the trust control layer and a simple *certificate linking* abstraction to organize logic content behind an application-specific trust API (§2). SAFE stores certificates in a shared key-value store (e.g., a DHT) indexed by parameterized string labels, which are hashed to yield self-certifying links. Participants may pass certificates by reference using their links, link related certificates to form DAGs, and retrieve certificates by their labels. A construction procedure integrates linking with the application's primitives for delegation and endorsement, naturally generating DAGs in which each certificate links to others needed to substantiate it (§3).

This approach enables us to factor DTM concerns out of the application and allows for richer control over trust choices without complicating the underlying logic language— on which the soundness of the trust system depends. At the same time, the logic language minimizes the need for custom software, certificate formats, and protocols for each application.

We used SAFE to implement trust kernels for example applications (§4). Experiments with a prototype (§5) show that trust scripts with certificate linking enable concise implementations. They can also reduce validation and proof costs by enabling a server to cache and index imported logic content and to select a tightly bounded subset of its cached content to use for each query or proof (*delegation-driven context pruning*,§6). A logic server may switch between cached sets to make trust decisions at high throughput. This combination of high-throughput logic servers and a scalable certificate repository enables SAFE applications to scale.

## 2. Background

SAFE is a data-centric platform for building multi-domain networked systems based on a *logical trust data plane*. This section summarizes security assumptions for the target applications, then explains the role of logical trust (§2.1) and the SAFE application model (§2.2).

**System model.** Participants of a multi-domain networked system are software programs controlled by principals (e.g., users or enterprises) and interacting via messages on the network, e.g., using some RPC protocol. Each participant trusts the others only to the extent provided by its *trust policy*, which governs its releases of information and its acceptance of incoming requests and assertions from other principals. Any participant may issue *assertions* about other parties and/or objects in the system. A trust policy may consider these statements by other parties in checking compliance.

**Security model: assumptions.** Each principal wields a keypair suitable to authenticate its communications. In particular, we presume in this paper that principals issue their statements as *certificates* signed by the issuer. Certificates have a timestamp and expiration time (TTL); clocks are synchronized tightly enough to use them.

**Trust model.** Programs and policy interpreters are under each principal's local control. We presume that principals use only trustworthy programs, and configure them with their keypairs: a program speaks with the keypair of a principal that operates it. A principal may configure its trust policies locally, embed them in the programs, or import them from another trusted party. Of course, a program may take inputs directly from the user to configure the policies, e.g., a user might use a program to create new objects and specify who can operate on them.

**Threat model.** The system protects against disruption by unfaithful or unauthorized parties injecting messages into the protocol. SAFE is concerned with the integrity of the trust policy and supporting infrastructure, which governs acceptance of assertions and requests from other parties. Participants may share policies and certificates voluntarily by passing them on the network; they are otherwise confidential, but their privacy is not paramount in this paper.

*Examples.* Familiar examples of this class of systems include critical services for a secure Internet, e.g., secure routing (e.g., Secure BGP or RPKI/BGPSEC), hierarchical naming (e.g., DNSSEC), and certification hierarchies (CA-PKI). The RPKI range delegations, BGP route advertisements, DNS name records, and PKI identity bindings are examples of "assertions" passed within signed certificates. Trust policies to validate these certificates are integral to these systems. BGPSEC routers verify that each step in a network route corresponds to a (transitive) advertisement of an IP prefix controlled by the origin, and DNSSEC clients accept name records only from servers that are authoritative for those names according to name delegations from a globally accepted root principal (a keypair). The standard CA-PKI trust policy accepts name-key bindings that are endorsed transitively from any locally accepted root Certifying Authority (CA).

These systems also illustrate local configuration of the trust policy. Each participant in RPKI/BGPSEC or DNSSEC has a local administrator who specifies *root anchors*—the public keys of the DNS root and of the Regional Internet Registries (RIRs) that allocate IP address space. Similarly, users of PKI software import lists of public keys for root CAs from parties they choose to trust; initial root CA lists are often embedded in the program itself (e.g., a Web browser).

## 2.1 Logical Trust

SAFE uses a logic to represent trust data exchanged on the network (in certificates). Logical trust provides a general and common language to express trust assertions, such as those as in the examples above. It also allows us to specify the trust policies as declarative logic rules that capture the necessary safety relationships among assertions in a concise, precise, and verifiable way. A participant can apply these logic rules recursively to create a logical proof that a given certificate delegation chain (or tree/DAG) is sound and compliant. For example, §4 describes SAFE script packages for secure routing and federated naming, modeling the structure of RPKI/BGPSEC and DNSSEC compactly using logic.

In logical trust systems, each participant uses a local interpreter and logical inference procedure (a *prover*) to check boolean-valued *guard* conditions for trust or access by issuing a logic query against a set of relevant statements (a *context*) that includes logical rules and facts asserted by other parties; a participant that checks a guard is an *authorizer*. In this way authorization is naturally end-to-end [19].

SAFE addresses support for logical trust as a systems problem, independent of the logic language in use. Its concepts (§3) apply to any declarative trust language that supports trust delegation among concretely named principals. However, we advocate use of pure Datalog, a well-studied declarative Prolog subset with a rigorous semantics [14], to represent trust content and trust policies. (See §8.1 for a rationale.) Our prototype uses an off-the-shelf open-source embedded interpreter called Styla as its Datalog prover.

**"Says" operator.** Styla's namespace feature offers a convenient syntax for the classical **says** operator introduced in BAN logic [12]. The **says** operator enables direct use of Datalog as a modal logic of belief and attribution (a trust logic), following Binder [16]. In our usage of Datalog as a trust logic, each atomic statement (atom) has a prefix parameter that represents a principal who **says** it (the *speaker*). The prover tracks the source (speaker) of each assertion or belief across inference. Policy rules consider the source in drawing inferences: a policy accepts and considers a statement $S$ only if it has a rule with a matching goal atom that confers trust in the speaker of $S$.

**Logical trust on the network.** All statements in a certificate must have a speaker that matches the certificate's issuer, or else the certificate is invalid. In this paper all principal IDs are hashes of the principal's public key, following SPKI/SDSI [17]. These IDs are used as constants in the logic: e.g., the value of a speaker is a principal ID, which is checked against the public key of the issuer who signed the certificate. The hash function—SHA-256 in our prototype—is presumed to be cryptographically secure. SAFE extends this idea to self-certifying object IDs: see §3 and Table 1.

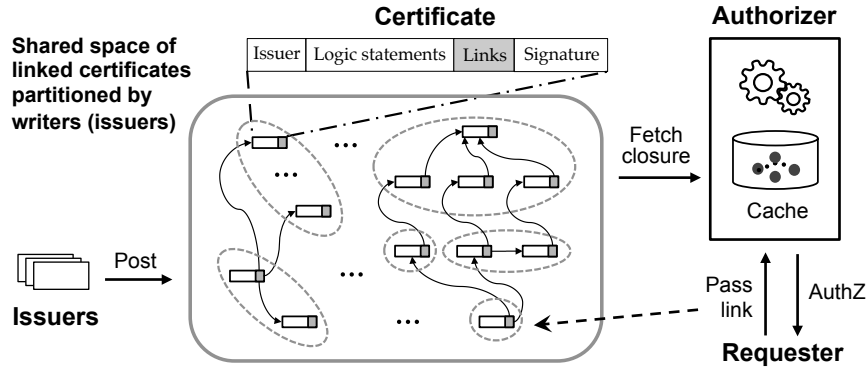

Figure 1: **Linked logic sets in a shared certificate store in a typical usage scenario. A requester passes a reference to a linked certificate DAG in an authenticated request. The authorizer uses the certificates to check that the requester's principalID is authorized for the request in compliance with the authorizer's trust policy.**

We do not assume any PKI infrastructure to distribute public keys: all common trust anchors and key endorsements are defined within the application. For example, any principal $P$ may issue any assertion about another principal $Q$ as a logic statement: is spoken by $P$ and names $Q$'s principal ID (key hash) as a parameter, and is passed in a certificate signed by $P$. A receiver of the statement validates its authenticity before accepting it into a proof context. Its prover then considers the statement according to the policy rules in the context. In this way logical trust extends conventional identity-based PKI security to incorporate rich statements about principals and their security attributes, and avoids the need for a global root anchor—although an application may choose to use one.

### 2.2 Building Systems with SAFE

An application of SAFE is a set of software programs that the principals deploy and use to act within the system.

**Development model.** Developers may write programs in any language, but use SAFE's scripting language (§3) to implement the trust kernel of the application as a module comprising a set of *trust scripts*. The trust module presents an application-specific API to the rest of the program, with a method for each trust-related event or action. The program invokes its trust API in response to incoming requests, user actions, or other events.

The trust scripts hide details of the trust implementation. They handle all details of generating, sharing, and processing certificates and their logic content, using the SAFE abstractions described in §3. Developers of trust scripts and policies must understand SAFE and its logic, but ordinary application developers see only the trust API, and users see only the application's external interface.

**Evolution.** Multi-domain networked systems evolve over time as principals join or leave and act according to their policies. Their trust scripts may generate and/or adopt new keypairs, issue certificates under their keypairs, withdraw or modify previously issued certificates, or retrieve certificates issued by other parties and interpret their content according to local trust policy rules. An invoked script may import new trust content (facts or policy rules) on the fly as data, e.g., from a file or from the network. In this way the policies and trust structure may evolve over time without modifying the software itself.

**Shared vocabulary.** To exchange trust data, principals must agree on a vocabulary of logic symbols (predicates and constants) and their meaning. A simple way to assure common conventions is for everyone to run the same (or compatible) trust scripts. Importantly, it is not dangerous for a principal to adopt incompatible or unfaithful trust scripts. The integrity of a SAFE system does not rely on the trust scripts, but on the soundness of the underlying logic. If a principal adopts malicious scripts it may harm itself and others who choose to trust it, but it cannot subvert the system.

**Trusted Computing Base.** Each principal's TCB includes its prover, script interpreter, and scripts, which are all under its direct control. Each principal must trust its own scripts, since they may issue certificates under the principal's keypair, and they also control how it validates, accepts, and interprets the certificates it receives from others. We emphasize that certificates never contain scripting, but only declarative logic content attributed to an authenticated speaker (with **says**).

### 3. Linked Logic Sets

The SAFE platform offers new abstractions and system support to simplify construction of multi-domain networked systems using trust logic, and to assure good performance for logic operations by careful management of proof contexts. Its novelty lies in a linked logic set abstraction, set-oriented scripting, and programmatic control of linking and

3                                                                                                                                                          *2016/11/9*

context assembly within the trust scripts. Certificates are shared by *posting* them to a shared certificate repository and passing them by reference. (See Figure 1.)

**Abstraction: sets, labels, and linking.** SAFE introduces an abstraction of *linked logic sets*. A logic set is an authenticated set of declarative trust logic statements gathered from one or more issuers. Every SAFE certificate encapsulates a logic set: the semantic content of a SAFE certificate is a set of logic statements authenticated to the certificate's issuer. SAFE certificates may also contain *links*, which are self-certifying references to other certificates, possibly from other issuers. The *closure* of a certificate is a logic set resulting from the union of the content of that certificate and of its linked descendants in the certificate DAG.

**Certificate storage and retrieval.** The issuer of a SAFE certificate names it with an arbitrary string *label* that is unique among all certificates issued by that principal. A link is a hash of the target certificate's label and its issuer's public key. Anyone can fetch a certificate given its link. Our prototype stores certificates indexed by their links in a shared key-value store. Prior research has shown how to implement key-value storage in a reliable and scalable way (e.g., as a DHT); §8 discusses issues for the certificate/store infrastructure.

**Scripting primitives for logic sets.** Trust scripts use builtin scripting primitives to create logic sets from parameterized templates, link them to other sets, materialize them as certificates, fetch and combine links and closures, and issue logic queries against unions of link closures.

Links serve as a general indexing technique for stored certificates. A script can synthesize a link given knowledge of the issuer's public key hash, its label conventions, and any parameters in the label. In effect, the model supports simple exact queries for certificates by synthesizing the link from a unique tuple of string values interpolated in a known label template. Some SAFE applications merely follow link trails left by issuers, but others synthesize links from shared conventions for label and logic templates embodied in the trust scripts.

### 3.1 Slang Scripting

We found it convenient to implement the scripting language with a separate interpreter, rather than embedding it in a general-purpose programming language. We refer to it as SAFE language or *slang*. We present slang details only as needed to illustrate the concepts of trust scripts.

Slang offers a simple functional model with recursion but no looping. The only data types are strings and named logic sets. It adds set-oriented scripting features to a logic interpreter core: functional rules that return values; logic sets as first-class entities; parameterized set constructor templates (`defcon` rules); template-based context assembly (`defguard` rules); programmable invocation of the prover to issue logic queries against assembled contexts; and an extensible library of useful builtins (see Table 2). Slang supports ordinary rule variables and also *environment variables* that are scoped to a slang interpreter thread, e.g., for command line arguments or script request parameters.

| Constant type | Description |
|---|---|
| `<predicate>` | A boolean-valued function, e.g., `evil(mallory)`. |
| `<principalID>` | A secure *principalID* is a hash of the principal's public key. |
| `<objectID>` | A *self-certifying identifer (scid)* is the principalID of the object's controlling authority concatenated with an identifier chosen by the authority (e.g., an RFC 4122 GUID). The builtin `rootID(scid)` returns a scid's principalID. |
| `<pathname>` | A sequence of components separated by /, optionally prefixed with the principalID of the naming root (like SPKI/SDSI names). Example: `bob:a/b/c`. |
| `<IP range>` | A standard IPv4 prefix/range descriptor, e.g., `ipv4"152.3.136.0/24"`. |
| `<value>` | Values are strings or numbers. |

Table 1: **Term constants for trust logic content, including conventions for self-certifying IDs and pathnames supported by slang to generate and query logic content. The meanings of constants are user-defined: if A has a policy rule with a goal atom that matches an assertion said by B, then A accepts B's meaning for constants in that assertion.**

| Function | Description |
|---|---|
| `post` | Post a set as a certificate and return its token. |
| `scid` | Generate a self-certifying identifier (scid) for a new object, for use as a term constant. |
| `principalID` | Return principal ID (a key hash) for a public key. |
| `rootID` | Given a scid, return the principal ID of its controlling principal. |
| `{...}` | Set template: return a set formed by interpolating slang variables in the templated logic. |
| `link` | In a set template, add a link to another set. |
| `label` | In a set template, identify the set by a string label. |
| `splitHead` | Return the first component of a pathname. |
| `splitTail` | Return pathname with head stripped off. |
| `tokenFromLabel` | Given a set label and principalID, return its token. |

Table 2: **Selected slang builtins for use by trust scripts.**

**Objects and scids.** Slang builtins support useful conventions for term constants in the underlying logic, including self-certifying pathnames IDs for principals and objects (Table 1). In particular, it supports *self-certifying identifiers (scids)* for "global" objects. The name of a global object is securely bound to some principal (its *controlling authority*) who controls the name. Scripts use a builtin function to generate scids, and the `rootID` builtin returns a scid's controlling principalID (Table 2). Self-certifying IDs ensure that parties have distinct names for their objects, and a malicious principal cannot "hijack" another's names: the BAN **controls** relation [12] is clearly defined.

### 3.2 Set Storage: Certificates and Tokens

If a script exports (posts) a locally constructed logic set, the slang runtime materializes it as a *SAFE certificate* authenticated to the issuer. It records the issuer's principalID

4                                                                                                                                       2016/11/9

($Self) as the speaker of each statement, and signs it under the keypair for $Self. The slang runtime writes a posted certificate to the shared certificate store (§5) indexed by a *token* that is globally unique and secure.

**Secure tokens.** A token acts as a secure link: a certificate is accessible to any client that knows its token, but only the issuer can post to the token or modify the certificate. SAFE generates a token for a posted logic set by hashing the principalID $Self with the set's string label, which must be unique among all sets controlled by the principal. Thus tokens are self-certifying and cryptographically unforgeable.

**ID sets and identity certificates.** By convention every principal posts an *identity/ID set* (equivalent to a self-signed identity certificate) containing its full public key under a null label, such that its token is the key hash (principalID) itself. SAFE uses the principalID directly as a token to fetch the ID set when the full key is needed to validate signatures of certificates issued by the principal. This convention enables use of compact key hashes as principal names.

Similarly, an object's controlling authority may post an ID set for the object under a token computed directly from the object ID (scid). An object's ID set is an object credential containing facts and/or policy rules about the object.

### 3.3 Set Linking

Secure tokens and the SafeSets certificate store enable participants to pass sets by reference, fetch them on demand, and cache them after validation (§5), indexed by their tokens. They also enable flexible *certificate linking*: if a posted set contains links, their targets are embedded in the certificate as tokens. Each token $t$ references a unique certificate $c_t$ and a logic set containing the logic content of $c_t$ and its closure—the content of the certificates linked transitively from $c_t$. The linked certificates form a DAG: fetches ignore any links that result in cycles.

Thus the construction of linked certificate DAGs is distributed across the participants who issue and receive them as credentials—sets/certificates containing endorsements and delegations. It is based on two principles: 1. The *explicit delegation rule* requires that each participant collects and stores its received credentials by linking them into *credential sets* that it maintains. 2. The *support rule* requires that the issuer of any new credential links it to any of its own credential sets that support its authority to issue the credential. If all issuers follow these rules then by induction the transitive closure of any credential contains the totality of upstream credentials needed to validate it—its *support set*.

Credential sets are also used at request time: a client passes a credential set token to a server (as $BearerRef) to substantiate an access request. These credential sets may contain a superset of what is required to support an endorsement or a request: the authorizer's prover extracts the relevant statements. A simple convention is that all participants maintain a *subject set* with a standard label for use to substantiate their requests and issued credentials.

### 3.4 Guards and Context Assembly

An authorizer's guard policies use logic rules to validate all credentials and delegations and verify that they are rooted in accepted trust anchors and compliant with the trust policy. Linking is also useful to assemble contexts for these authorization queries. Different trust decisions by the same principal may reason from different contexts. To this end, the slang runtime assembles the proof context for a query from one or more sets, each referenced by a link in a set template in the guard policy's defguard rule. With this structure an authorizer can obtain all credentials needed for a query by linking tokens into the proof context, then "pulling" and caching their closures.

**Summary.** SAFE generalizes certificate chaining to support DAGs. The DAGs are collaboratively editable: each certificate is controlled by its issuer, and changes by its issuer become visible in any sets that link to it. The links and common repository enable a guard policy to identify and obtain the certificates that are relevant to a particular trust query, and exclude others, keeping the context small for efficient queries (*delegation-driven context pruning*). While logic set linking and certificate chaining can be beneficial in other settings as well, the combination of linked certificate DAGs and trust logic is uniquely powerful. Trust scripts express an application's linking structure concisely, integrated with its use of the logic.

## 4. Example Applications

We contend that SAFE's abstraction of linked logic sets enables us to specify and implement multi-domain systems in a compact, flexible, and efficient way on a common platform. To evaluate SAFE we implemented some exemplary applications. These are script packages of common elements that may be combined in larger applications. Each of them illustrates different aspects of trust solutions using logic, scripting, and linking. Table 3 summarizes the applications and their complexity.

**STRONG** implements secure symbolic pathnames and nested groups with roles and delegatable membership. Principals issue symbolic names for principals or objects, including naming contexts of other principals. Pathname lookup is relative to a specified naming root anchor, which may be a shared naming root as in DNSSEC. Principals create groups and grant membership in them to principals or other groups, with optional named roles. Principals may also define groups and delegate nested group memberships within a federated space of named groups, as in SPKI/SDSI and SFS [22]. Groups may be used as roles or tags, e.g., in object ACLs. The combination of groups and names is equivalent to SPKI/SDSI [15, 17], with some differences: e.g., SPKI/SDSI groups are names, while STRONG groups are objects, which may have names). STRONG offers key elements of Amazon's Identity and Access Management (AWS-IAM [4]), and extends it to a multi-domain system. It adds a wildcard fea-

5                                                                                                                                                           *2016/11/9*

| Application | #Predicates | #Policy rules | #Assertion types | #Guard APIs | #LoC | #LoL | Description |
|---|---|---|---|---|---|---|---|
| **STRONG: naming & groups** | 14 | 10 | 17 | 5 | 201 | 79 | Groups with nesting, roles, objects and ACLs, nested groups, secure symbolic pathnames |
| **GENI** | 26 | 28 | 24 | 3 | 293 | 143 | Trust core for federated IaaS (NSF GENI [11]). |
| **Routing** | 5 | 5 | 11 | 2 | 95 | 41 | Secure prefix delegation (e.g., RPKI) and route advertisements (e.g., BGPSEC) |
| **Attestation** | 11 | 4 | 8 | 2 | 103 | 43 | Secure code identity, code attestation chains, image properties, attestation-based access control for principal ID or IP address. |

Table 3: **Example applications built on SAFE and their complexity: trust script APIs, lines of scripting code, and the number of lines of logic templates (LoL).**

ture from AWS-IAM to grant access for all objects with names matching a pathname prefix.

**GENI** is the trust core of NSF GENI [9], a federated Infrastructure-as-a-Service (IaaS) testbed. GENI's trust system extends the Slice Facility Architecture of PlanetLab (see [35]) as described in [11]. GENI serves as a full-featured network trust example that includes global objects (projects and slices: [11], §2.1) with hybrid capability-based access control, multiple object authorities ([11], §2.2), and a root trust anchor that endorses the authorities and member sites for a federation.

**Routing** implements secure IPv4 prefix delegation and routing: it checks the security of route advertisements (following an example in [38]) and prefix ownership. We added builtins to slang and Styla to manipulate IPv4 prefix types and verify membership and containment.

**Attestation** implements access control based on attestation of the code running on a client and properties of the code. A guard checks whether the code (image) running on a client can access a given object, based on an object ACL that lists properties of permitted image. The trust policy permits the access if the client is attested (e.g., by a trusted cloud provider) to be running a given image, and the image is endorsed as having a permitted property (e.g., by a trusted certification authority).

Figure 2 illustrates how the trust scripts for these applications link sets to according to the delegation patterns.

Datalog is sufficiently powerful to represent common features for trust and access control used in these examples. User-defined predicates may represent properties, attributes, roles, relationships, rights, powers, or permissions. An application can use its vocabulary of predicates and constants to represent common forms of restricted delegation: names, capabilities, endorsement/attestation chains, address range allocation, route advertisements. Each delegation is constrained by a predicate and its parameters, including the receiving principal (delegate) and other qualifiers such as an object name (e.g., *"Alice owns file F"*). Thus simple logic assertions can capture hierarchical naming, nested groups, named roles and other attribute assertions, ACLs, capabilities, and authority over ranges (e.g., IP prefixes). Declarative policy rules capture the conditions to validate these assertions. Conjunctive policy rules permit reasoning from multiple attributes of a principal or object. Policy rules are mobile: they may be passed in certificates.

**Example: capabilities.** Both STRONG and GENI support capabilities—transitive delegations for rights over objects, e.g., for membership in groups and projects and access to slices. Capabilities are represented as logic certificates that grant a named principal a specific named right over an object named by an ID. The named holder may in turn delegate the capability to others, in compliance with other applicable policies of the object's controlling authority. Delegations may refine the delegated rights, or restrict further delegation.

Listing 1: Policy rule for capability-based access control.

```
defcon capabilityRule() :- {
    cap(?Subject, ?Object, ?Priv, ?Delegatable) :-
        ?Delegator: delegateCap(?Subject, ?Object, ?Priv, ?Delegatable),
        cap(?Delegator, ?Object, ?Priv, true).
}.
```

The meaning of capability-based access control is easily captured in a recursive logic policy rule (Listing 1). It says that a delegate (`Subject`, a principalID) has a capability for privilege `Priv` over `Object` (a scid) if another principal (`Delegator`) has issued the capability, and the `Delegator` itself possesses the capability and is empowered to delegate it. The rule supports confinement: each `Delegator` specifies whether its delegate may pass the capability on to others.

**Set linking for capabilities.** In the scripts, a principal links its capabilities for an object into a *capability set* for that object. The capability set is an example of a credential set. If it delegates a capability, it issues a new set with a `cap` statement, links it to its own capability set and subject set, and passes its token to the delegate. Transitively, any capability token links the full delegation chain to validate it according to the policy rule.

**Pathname resolution** in STRONG is enabled by both scripting and set linking. Although logic can represent the structure of the name space, logic cannot resolve names: the entire context for a logic query is present at query time, so it does not block. Therefore, logic itself cannot fetch the name entries for each component in sequence. Instead, a logic script fetches the name set for each component by synthesizing its token from a standard set label (parameterized with the name) and the principalID of its parent domain. The script uses a logic query to extract matching name assertions from a name set to target the next fetch. Once the name sets for all components are present, the script invokes logic to validate the path end-to-end. CoDoNS [31] uses a DHT to resolve names in a similar way, and STRONG's validation step is exactly the *certified evaluation* of SD3 [21]. These works explore secure naming in the context of DNSSEC; STRONG generalizes these approaches to multi-rooted linked local names as in SPKI/SDSI.

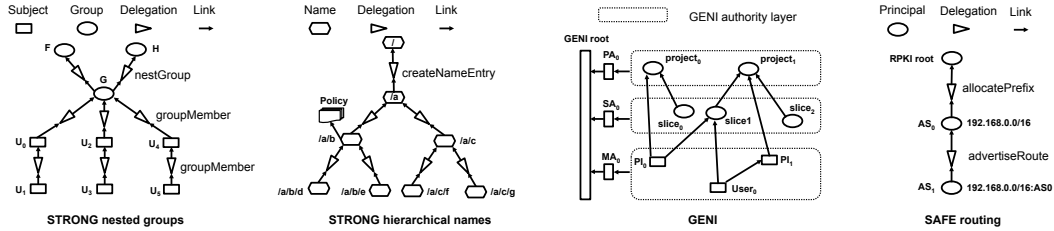

Figure 2: **Set linking examples.** For nested groups, links traverse group delegation and membership delegation backwards; for hierarchical names, links traverse delegation of sub-name spaces and policies on the object hierarchy; in GENI, links traverse delegation of projects and slices, and the endorsement by each layer's authority. Secure routing links to the IP prefix allocations and route advertisements needed to substantiate each route update.

## 5. Implementation

SAFE is implemented in about 9000 lines of Scala code, with Styla as an open-source Datalog engine. SAFE runs as an interpreter with one or more slang programs (logic scripts) loaded into it. The behavior of a script is determined in part by the logic content passed to it: participants may add new rules to evolve the policies or tailor them to local needs without changing the scripts. Scripts are composable: it is easy to add code to customize the local behavior. Changing the scripts does not affect other software and state at the site. The interpreter is stateless (except for the caches): it may restart and/or reload scripts at any time.

**Certificate handling.** Applications invoke the script entry points to handle trust-related events, e.g., checking access or issuing endorsements, delegations, or policy rules. The runtime system handles generation of certificates (in a native SAFE format) and tokens and related cryptographic operations automatically and transparently. It also handles post/fetch operations on the certificate store: it performs cycle-safe recursive fetches and parallelizes them using a thread pool (futures) to reduce latency.

The certificate store is a generic key-value store (we use Riak [32] in our experiments) with added write access checks (below): it verifies that the issuer of a post on a token uses a keypair whose public key yields the token when hashed (SHA-256) with the label. We refer to the store as "SafeSets".

**Set caches.** Logic sets created in slang or fetched from the store are cached in an in-memory *set cache*; fetch checks the cache for each subset token encountered during a recursive fetch. When a context is assembled for dispatch to the inference engine, SAFE *renders* it to an internal context format required by the Styla engine. We supplemented the context format and access procedures in Styla to add a *secondary index* to access statements by head predicate and first parameter, which we found important for to speed up the search for statements matching a goal in processing the delegation chains that are common in trust logic (§6). Contexts are cached in a *context cache*.

The caches store only fresh authenticated logic content, to assure the validity of logic content passed to the proof engine. Logic content expires from the cache at the expiration time of the containing certificate. A cached context expires at the earliest expiration time of any statement it contains. Currently there is no result caching.

**Deployment.** SAFE may run in the same JVM as the application, e.g., to implement command-line tools. A SAFE instance can also be configured to export script entry points through a generic REST API. This deployment model facilitates cross-language interoperability: an application server (e.g., a GENI control server) invokes its scripts over a secure channel to a separate SAFE process, which holds its signing key. (See Figure 3). The app server may use any language, as long as it can make REST calls. When a request enters the app server, it invokes SAFE to evaluate a matching guard, passing a list of string-valued parameters named in the request, including the subject's principalID and a credential token (`$BearerRef`). If the guard returns false, the server denies the operation.

The explicit delegation rule requires that issuers must pass tokens to notify a recipient of an issued credential; the recipient then invokes a script to add the link to a credential set. Currently, this passing of tokens must occur in an application-layer protocol. Tokens are base64-encoded strings that may be passed via web or e-mail.

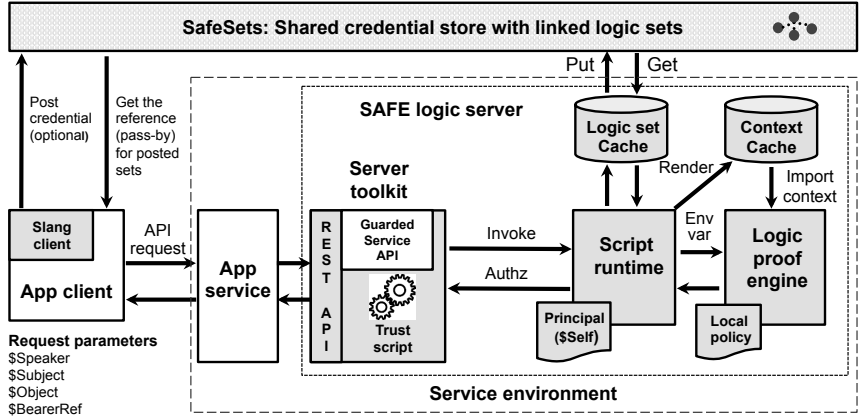

Figure 3: **Server access control using** SAFE. A SAFE instance runs as a separate process with loaded trust scripts (in slang) and the principal's signing key (`$Self`) and local policy rules. The server application invokes its scripts through a REST API. Credentials are passed as references to signed logic sets in a shared store (SafeSets). The SAFE process fetches sets on demand, validates them, and caches their logic content for use by the prover. Shaded components are application independent.

**SafeSets.** SAFE can also act as an access-control proxy (a shim) in front of the app service. SafeSets is itself implemented as a SAFE instance running as an HTTP/REST interception proxy to an unmodified Riak key/value store [32]. A slang program in the shim guards access for Riak *put* operations (*post*). It accesses operands of the request using slang builtins for SAFE's certificate format. It validates that the post token (the Riak key) is derived from a hash of the signed string label and the requester's public key.

The issuer's scripts control the lifetime of a certificate by setting its expiration date as a meta-predicate in the constructor. In addition, the issuer/owner may revoke a certificate early or reissue an updated certificate under the same token.

## 6. Evaluation

We run SAFE and application trust scripts (§4) on a cluster with multiple SAFE instances representing multiple principals in a multi-domain system, and subjecting them to request workloads through the REST API. For this evaluation the REST calls originate from a load generator harness. It generates script requests according to a synthetic mix, and maintains lists of principals (keypairs) and objects in various states to select from (randomly) as operands. We use this setup to evaluate four central points:

1. *Scripted set linking enables fast automated retrieval of the relevant logic content for each query or authorization decision.* The application trust scripts handle all linking and access checks required for their policies. The linked certificate DAGs flow naturally from delegation patterns in the request mix.

2. *Caching reduces the costs to pull and validate signed sets.* Commonly accessed logic content tends to reside in caches. In GENI the cached portions of the DAGs include the base trust structure, with endorsements of authority servers from the federation root, and packages of access policy rules from the qualified authority servers. In STRONG it includes portions of pathname trees close to their roots. The cache also captures repeated accesses.

3. *Scripted linking enables lightweight queries and trade-offs between policy flexibility and cost.* A core design principle of SAFE is to limit context size by including only the logic relevant to each guard query. Linked certificate DAGs can substantially reduce query cost by focusing the prover on the statements of interest: *delegation-driven context pruning*. In practice, script developers have a continuum of linking choices balancing various goals, but these choices are made in advance of query time, at delegation time—they are in essence pre-materialized views over the global set of credentials.

4. *Programmable linked contexts enable practical high-throughput logic servers.* Linking in guards enables a logic server to switch between contexts for logic queries. Each context includes linked closures selected for the needs of the specific query; linked pruning excludes superfluous statements from the context, reducing proof costs. The secondary index in the prover (§5) enables it to assemble proofs in time linear with the length of delegation chains, which occur frequently in trust logic. Some policies show nonlinear proof costs when the prover is forced to explore each branch of a disjunction (e.g., an ACL), but this cost scales with the complexity of the authorizer's policy.

Each SAFE instance runs on a four-core KVM (Intel Xeon CPU E5520 @ 2.27GHz) with 12 GB of RAM and 1Gb/s Ethernet. The SafeSets cluster is five similar VMs running Riak 2.1.4 [32] with a replication degree $N = 3$ and $R = 1$, $W = 3$. Each logic set is materialized as a certificate with a 2048-bit RSA signature; their logic payloads range from 467-840 Unicode characters. Tokens and principal IDs are self-certifying 256-bit SHA hashes (32-byte base64-encoded).

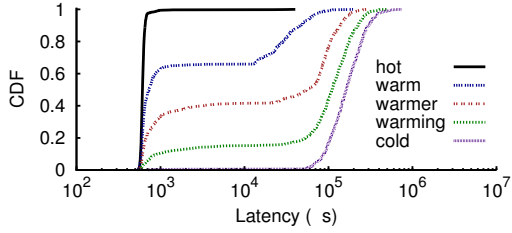
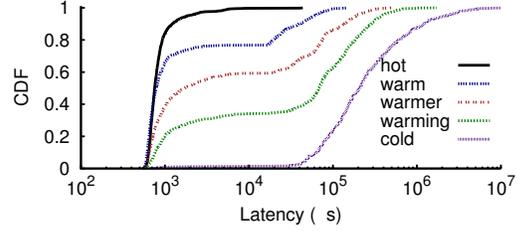

(a) **Direct linking.** With a hot cache this mix sustains 2754 authz-ops/sec under load with median latency of 670 $\mu$s.

(b) **Coarse-grained linking.** Poor context pruning leads to long tail latency. With a hot cache this mix sustains 2538 ops/sec with median latency 731 $\mu$s.

Figure 4: **End-to-end latency profile of queries for a delegation-rich GENI scenario (query-only, concurrency level 12).**

|  | Direct | Coarse |
|---|---|---|
| #sets in context (95%) | 36 | 192 |
| #stmts in context (95%) | 54 | 245 |
| Fetch closure (95%) | 184 ms | 298 ms |

Table 4: **Effect of linking granularity on the latency of closure fetches for the example of Figure 4.** Latency is measured on a SAFE instance with a warm cache (round 50). Direct linking reduces closure sizes (via better context pruning) and tail latency for closure fetches. SAFE's parallel closure fetches mitigate the higher tail latency for coarse-grained linking.

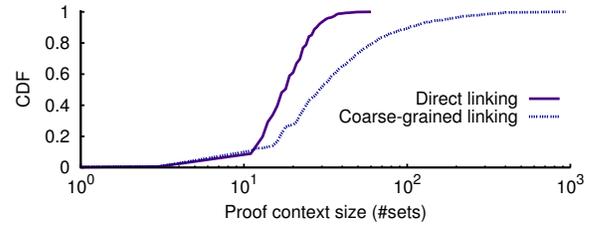

Figure 5: **Context size (number of sets) for the example of Figure 4.** Coarse-grained linking using subject sets is simple, but it produces large contexts in this delegation-rich scenario. Logic scripts with direct linking maintain multiple credential sets and exclude irrelevant sets (delegation-driven context pruning).

**Authorization cost with set linking.** We first explore the cost for a participant to check trust compliance in a large system—one with a large number of principals and credentials—under alternative set linking structures. We use a GENI mix to check access to slice objects for user requests. The policy checks that a slice is endorsed by an authority, that it is bound to a project endorsed by an authority, that the authorities are endorsed by the federation, and that the user has received delegated capabilities for the project and slice by a chain rooted in their owners. It checks against a pre-constituted certificate set with 1M users, 100K project groups and slice objects, and 500K random delegations: the delegation chains are longer than is likely to occur in practice.

Figure 4 shows the latency profile for processing and fetch costs for a server to check and approve random allowable requests. The experiments start with cold cache, and random requests are issued in rounds of 1000: the graphs plot latencies for selected rounds to show the effect of cache warming. As the caches warm, latencies drop and fewer requests show long latencies due to misses and context assembly. The caches are sufficiently large to avoid eviction for this workload. The hot results show the in-server processing costs. All other costs are driven by certificate fetches: with a scalable K/V deployment this system could scale to a large number of issuers and authorizers.

**Effect of set linking choices.** Figure 4 also shows that linking conventions in the scripts can have a substantial impact on context size and proof costs. Linking is *coarse-grained* (right graph) when it follows the simple convention in §3.3: each principal maintains a single credential set (subject set) for received delegations, and links it as the support set for all issued delegations. Linking is *direct* (left graph) when a principal's subject set includes only its identity credentials (e.g., a principal's attribute endorsements); received delegations are maintained in separate credential sets, and each request or delegation links only to the subject set and selected credential sets needed to substantiate it. Direct linking requires a 19-line change in the GENI scripts. It yields a substantial improvement in delegation-rich scenarios: with coarse-grained linking, the delegator links each delegation to its subject set, which links to many irrelevant received delegations. Table 4 and Figure 5 show that direct linking prunes context more effectively, reducing fetch costs, improving cache effectiveness (as shown by comparing the warm Figure 4 curves), and reducing query costs (as shown by comparing the Figure 4 hot curves).

**Cache updates.** SAFE uses a simple policy to handle updates to credential sets in the cache. Figure 6 continues the GENI scenario with 40K principals, but adds new delegations to the query mix. These results use direct linking for delegations, but the client passes its subject set as the $BearerRef to substantiate a request. When the cache is hot, an update-then-query sequence (e.g., new delegateS-

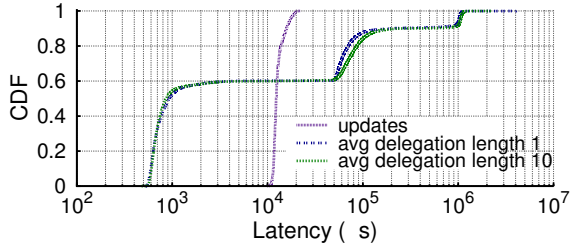

Figure 6: **Latency profile under a mix of queries and certificate updates on a cluster. The workload is under the GENI scenario but adds random update-then-query sequences (new projects/slices and delegations) to the mix. The target server serves queries with a hot cache. Certificate updates to SafeSets are sent to other servers in the cluster, which lead to failed inference on the target server that need the new delegations. The update-then-query sequences constitute 40% operations in the mix, among which 25% updates are delegations each of which immediately follows a predecessor update issued from the same harness thread and is made to the same subject. The results show the latency of each query on the target server and the latency of each update collected from the cluster. The query latency results show a spike for the resulting failed inference and invalidation/refresh of the cached context and its sets under updates and another spike for the delayed cache refreshes.**

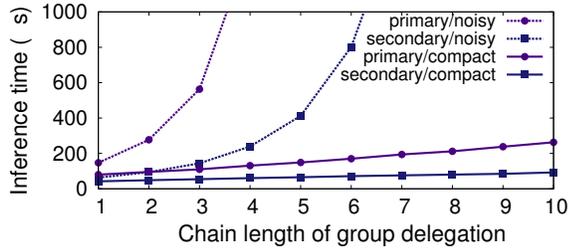

Figure 7: **Comparison of inference cost for group delegation between noisy and compact contexts, with primary and secondary indexing schemes. The baseline delegation chain contains 1 group membership and 5 membership delegations.**

lice, then update capability set, then sliceOperation such as createSliver) results in a failed authorization, because the cached subject set does not reflect the new object and delegation. In this case, SAFE invalidates and refreshes all sets in the context, and then repeats the guard query. To avoid a denial of service attack it adds a variable throttle delay of up to one second before the cached context or sets can be refreshed again (in addition to an optional limit on the number of statements or sets in a context). Figure 6 shows the latency of the update operations and also the failed queries and context reload.

**Context pruning and indexing.** Certificate linking and delegation-driven proving can also improve the efficiency of logic inference substantially. We show this effect for nested groups (Figure 7) and named objects (Figure 8) in the STRONG group service and naming service respectively. The structure of group delegations creates a hazard for a prover trying to answer an access query: *is principal $P$ a member of group $G$?* If membership in $G$ is delegated to multiple subgroups, recursively, then the prover may explore the subgroup closure searching for a subgroup that includes $P$. In SAFE, these superfluous delegations are pruned, because they are not reachable in the link closure of principal $P$'s BearerRef. That is $P$ does not link any of its credential sets to them, because there is no associated delegation of interest to $P$.

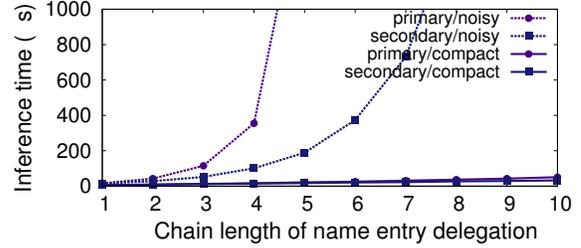

Figure 8: **Comparison of inference cost for name entry delegation, between noisy and compact contexts, with primary and secondary indexing schemes.**

We show the impact of linked pruning on the inference using noisy contexts with superfluous delegations. The linking patterns in the scripts prune these superfluous delegations, so a SAFE prover would not see them in real operation. In a noisy context, we inject the distracting delegations as a binary tree and set the tree height to the length of the delegation chain (given by the x-axis). We conducted similar experiments with the same methodology for subgroup delegation (Figure 7) in the group service and for sub-namespace delegation (Figure 8) in the name service; in the naming example, to distract the inference we add a name entry delegation tree under the root naming object that has been attached access policies as ACLs. The query checks whether a target object is under the root naming object; the height of the noisy delegation tree is set equal to the length of the pathname (given by the x-axis).

For both examples, the noisy contexts lead to exponential proof costs. In contrast, *the proof cost is always linear in the length of the delegation chain for the pruned context in conjunction with the secondary index*. In the group example, the prover explores it by applying group delegation rules backwards recursively from the root, and at each step it knows the containing group that it is looking for (initially it is $G$). The primary index indexes on the goal's predicate name and argument count, but the secondary index includes this first argument. It takes time linear in the number of facts to build the secondary index for a context, but once it is present the desired fact is extracted from the context in constant time. The trust scripts choose the order of parameters to ground facts representing delegations to enable effective indexing with the secondary index.

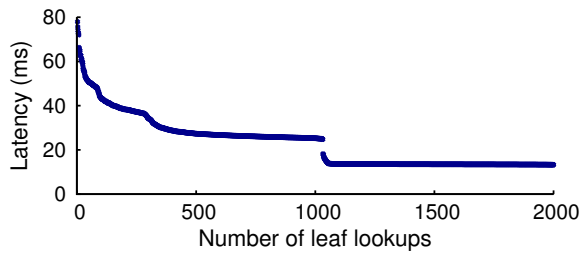

Figure 9: **Latency of bulk name lookups with caching. As the client's set cache warms, lookups speed up as the interior of the naming tree is cached. On a hot cache, resolution of a leaf name takes 13.3 ms (95%).**

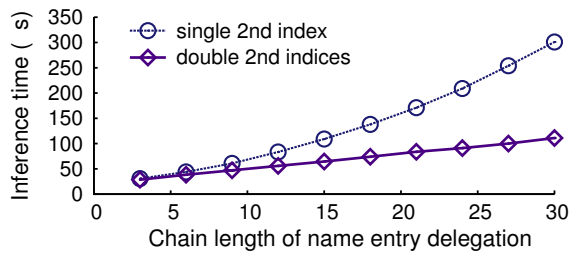

Figure 10: **Cost of access compliance checks in an IAM-like environment enabled by STRONG. The subject of a query is an immediate member of a group that is authorized to access a directory containing the target object. We vary the chain length of name entry delegation between the containing directory and the target object and compare the inference time with single secondary index and with double secondary indices on name delegation.**

| | Operation | latency |
|---|---|---|
| **Post** | sign/issue | 8.0 ms |
| | post data | 3.9 ms |
| | post validation | 322 $\mu$s |
| | script processing | 13.4 ms |
| **Fetch set** | fetch | 3.2 ms |
| | verify signature | 294 $\mu$s |
| | parsing | 3.2 ms |
| Name validation (CE) | | 134 $\mu$s |
| Lookup processing (one hop) | | 30 $\mu$s |

Table 5: **Breakdown of operation latency (95%) in naming service, to issue authenticated names (post) and resolve a pathname component (fetch/verify/parse). "Script processing" includes all other costs.**

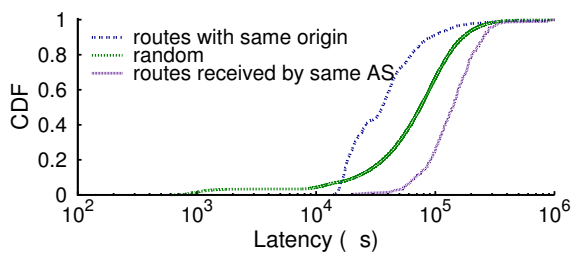

Figure 11: **Latency of routing verification under three query workload models: routes with the same origins, routes randomly selected, and routes received by the same ASes.**

**Latency and caching in the naming service.** Pathname lookups are a different case because the client trust script must parse the pathname, extract a name fact (via a logic query) from a logic set at each step, and synthesize a token to fetch the logic set for the next step, followed by a final query to validate the pathname. It also demonstrates the cost of script processing and the benefits of caching when the client resolves many names. We build a synthetic naming tree with 1024 pathnames (height 5, branching factor 4). The test harness sends a sequence of queries on the 1024 leaf names. Figure 9 shows the latencies of these sequential lookups: queries beyond 1024 are repeats that resolve from the cache, with all costs due to processing the scripts and logic.

This example also illustrates the benefits of combining logic with a scripting layer: no name server is needed, since all name entries reside in the certificate store and are authenticated by their issuers. Table 5 shows the breakdown of processing costs to create names (via post) and perform lookups. The post results show the costs to sign/issue certificates and post them to the store, and the scripting costs.

**Multi-indexed facts.** In some cases, the desired order of parameters is not unique due to different types of queries. For example, in an AWS-IAM environment powered by STRONG, a guard checks whether a subject can have access to a named object. Clients attach policies (e.g., ACLs) to an owned directory which apply to all objects with names matching the pathname prefix of the directory. The service may need to search up the name delegation chain for certified evaluation of object pathnames (to validate resolved pathnames) or down the chain for prefix access checks (to check if the target object is under an accessible directory). The STRONG trust script issues two facts in the logic set for each delegation, using different predicates: they have the same meaning, but different parameter orders. Each policy rule uses the predicate with the most effective indexing for its query. Figure 10 shows the difference of inference time with single secondary index and with double secondary indices.

**Routing and IP delegation.** We examine the efficiency of queries in the **Routing** application under a synthetic scenario (Figure 11). AS principals exchange information on routing and IP prefix assignment/delegation in the same way as in BGP and RPKI. A guard checks whether an advertised path is valid by verifying route delegation at each hop and IP prefix ownership of the origin. We split the IP address space into 32K prefixes and delegate those prefixes from an (RPKI) trust anchor to 32K ASes. The branching factor of the delegation tree is 8 and the depth is 5. We generated a random network topology of those ASes, and ASes choose the shortest path to propagate routes. The average path length is 6.9.

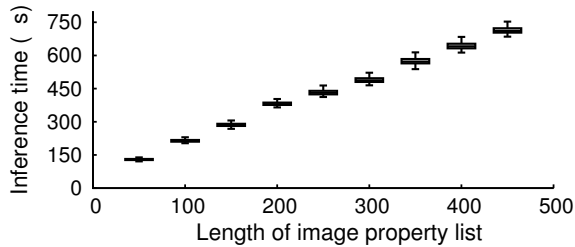

Figure 12: **Inference time as a function of image property list. SAFE inference engine iterates through all properties associated with the image. The inference cost increases almost linearly with the length of image property list. The length of ACL list of the target object is 2000.**

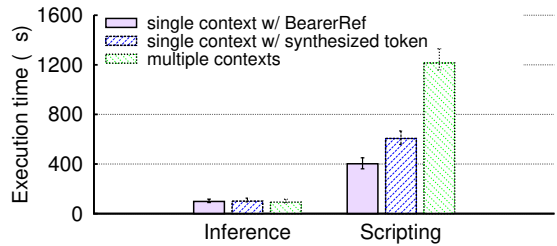

Figure 13: **Comparison of execution time of logical inference and end-to-end slang processing under three common slang programming patterns.**

Figure 11 shows the latency distribution for a single SAFE engine to bulk-validate all routes with the same origin, routes received by a given AS, and a random set of routes. This example illustrates a limitation of SAFE's linking and "pull-based" approach, which is not suitable in all cases. Pulling and caching help to validate random routes and for routes with the same origin. However, in BGPSEC each node (e.g., a router or host) must validate the routes that it receives, and each route advertisement is specific to the advertiser and the destination prefix. Thus a node considers each certificate exactly once, and its set cache yields no value. Moreover, "pulling" certificates on demand incurs fetch costs, while the conventional approach of "pushing" supporting certificates with each advertisement has the desired effect of leaving each node with exactly the certificates it needs, just when it needs them. Each certificate is needed by all successors in its path, but never by any other node. This is the worst case for SAFE's pull-based approach.

**Attestation-based access control.** In attestation-based access control, a guard checks whether the code (image) running on a client can access a given object, based on an object ACL that lists properties of permitted image. The trust policy permits the access if the client is attested (e.g., by a trusted cloud provider) to be running a given image, and the image is endorsed as having a permitted property (e.g., by a trusted certification authority). In this experiment, the target object has 2000 entries of image properties in its ACL and the number of properties the image of the requesting client has varies from 50 to 450. Figure 12 shows performance for guard queries for attestation-based access control. The inference engine iterates through all image properties and the secondary index ensures checking whether a given property is on the ACL list completes in constant time. As a result, the inference time increases almost linearly to the length of the image property list.

**Processing cost of Slang programming patterns.** We use attestation-based access control to explore the cost of scripting in difference programming patterns. In this experiment, the requesting client has 5 properties on its image and the target object has 7 entries of image properties in its ACL. We wrote slang script for the guard with three different patterns: single context with BearerRef, single context with synthesized context token, and multiple contexts. As shown in Figure 13, the inference costs among these three patterns are close, because the underlying logic queries and statements are the same. The script using single context with BearerRef is the most efficient as the server need neither synthesize tokens nor assemble multiple proof sub-contexts. At the same time, the scripting cost is still moderate with each pattern. In practice, a developer can balance between programming efficiency and scripting overhead to choose a suitable pattern.

## 7. Related Work

**Trust logic.** SAFE builds on trust logic, but its contributions (scripted linking) are independent of the trust logic in use. SAFE could use any declarative language that supports trust delegation among concretely named principals: §8.1 discusses background and alternatives. We advocate use of Datalog as the most powerful logic that is tractable. One role of this paper is to demonstrate its practicality for a range of examples. Although we do not expect existing systems (e.g., DNSSEC or BGPSEC) to be reimplemented using logic, the examples show that logic can be a powerful basis for future multi-domain systems.

Our SAFE prototype uses a Datalog-with-says trust logic equivalent to Binder [16]. SD3 [21] and SENDLog [2, 38] also use Datalog-with-says: our naming and routing applications are based on examples developed in these prior works. These languages add new features for distributed coordination within the logic. In contrast, SAFE uses pure Datalog within a larger framework to organize and share logic content in a networked system based on the novel abstraction of linked logic sets.

**Credential discovery and PCA.** Logical trust systems enable flexible declarative policies, but do not in themselves address the credential discovery problem—identifying and obtaining the credentials needed by a policy in a distributed setting. Many previous works simply assume that a caller identifies the correct credentials and passes them by value in each request, which may force the receiver/authorizer to pro-

12                                                                                                                           2016/11/9

cess them even if it has already received them for a previous request. SAFE avoids this cost using caching and pass-by-reference of linked certificates named by secure links. Proof-Carrying Authorization (PCA [7]) goes further in requiring the caller to complete the proof itself. We show that linked certificates with delegation-based context pruning can yield proof costs low enough to avoid this burden.

In other approaches to credential discovery the prover issues distributed queries as it runs (e.g., [6, 15, 29]). In contrast, SAFE indexes and retrieves certificates by hashed parameterized string labels in scripting outside the logic layer, enabling the prover to run without blocking. It supports retrieval of linked DAG closures for explicit delegations, and precise and flexible queries from the shared store when label parameters are known. We show that these features address credential discovery in a simple and uniform way, clearing the way for practical use of trust logic for many applications.

**Certificate storage in a DHT.** SAFE facilitates certificate sharing via a key-value store model (e.g., a DHT), which can be highly available and scalable. The early X.500 model proposed a distributed certificate repository, as discussed by the SPKI/SDSI authors [17], who judge it to be impractical. A key difference is that SAFE certificates are indexed by hashed links: like SPKI/SDSI they do not require coordinated global principal names, which are instead built at a higher layer. Many DHT applications have stored self-certifying content by hashes in a similar way (see [13]).

Conchord [3] introduced the use of a DHT as a shared store for logical certificates. Conchord preserves closure under inference for stored certificates: insertions trigger recursive rounds of retrievals and new insertions of newly derivable facts. Conchord's insertion algorithm is specific to the limited representational power of SDSI, and does not generalize to Datalog, which is a more powerful logic. CERTDIST [34, 35] follows the query model of certificate discovery; it uses a DHT as a cache, but leaves programmers to decide how to index certificates, e.g., by hashing selected fields to assign the keys. In contrast, SAFE offers a simple and uniform linking model outside of the logic layer, which enables simple indexing of the DHT and local caches, and serves the needs of key applications. It also generalizes certificate linking within the DHT in a novel way. Note that SAFE's linking at the set/certificate level is distinct from SDSI's "linked local names": these exist at the logic level and are subsumed by Datalog [26], which SAFE also uses.

CoDoNS [31] proposes a DHT to store signed DNSSEC name records, and shows that it can improve DNS lookup behavior. We show how to use this approach within SAFE to support multi-rooted hierarchical naming. SAFE's scripting layer supports name lookup by indexing each name component with a link (§4), followed by an end-to-end logical validation check similar to the *certified evaluation* of SD3 [21].

## 8. Discussion

### 8.1 Toward a Standard Trust Logic

Many works have developed languages for managing trust and access control. Any such language reduces to a logic if it is rigorously defined. For example, it is now understood that XACML access control—now common in industry—is equivalent in power to Datalog. Its uptake suggests that the expressive power of logic has value in enterprise information systems and other settings in which XACML is used.

Logical trust systems are often called Logic-Based Authorization (LBA, e.g., Nexus/NAL [33, 36]); Proof-Carrying Authorization (PCA [7]) is also logic-based. LBA is a powerful building block for secure networked systems. Applications of LBA include tag-based file access [30], building controls [5], software trust with logical attestation [33], and Internet infrastructures including DNSSEC [21] and secure routing [2].

We advocate Datalog-with-says as a practical standard for network trust logic. Off-the-shelf logic engines (e.g., the Styla engine used in our prototype) offer a standard syntax for the trust logic **says** operator within Datalog. Briefly, any predicate symbol may be qualified with a ":" operator and a *namespace* value (a string). The namespace acts as syntactic sugar for a "hidden" first (prefix) parameter of the predicate. Li [25] reduces Datalog-with-says to Datalog using an implicit prefix parameter for each predicate and atom in a similar way.

In this way Datalog can represent precise restricted delegations using **says** and inference. In particular, a statement about a principal naturally represents a delegation or endorsement that is restricted by the speaker and predicate. An authorizer accepts the delegation only if it has a matching rule in its trust policy that trusts the speaker as a source for the specific predicate and parameters. The speaker of a goal atom in a rule may be a variable; quantifying over principals in this way enables *attribute-based delegation* [27], a useful feature that bases acceptance of a statement on what other principals say about its speaker.

Classical authorization logic (ABLP [1, 23] and its successors [19, 33]) also rely on **says** as a cornerstone, but introduce a second-order operator (**speaksFor**) as the basis for delegation. These logics are expressive, with a rich notion of compound principals, but are intractable and therefore unappealing for practical use. In contrast, Datalog can represent precise restricted delegations using **says** and inference, without **speaksFor**.

Since Datalog with ordered domains is P-complete [20], Datalog-with-says is the maximally expressive tractable trust logic. Some early trust logics are less powerful and lack essential features. SPKI/SDSI [15, 17] and role-based trust (RT0 [28]) can represent rich naming and access control, but are weaker than Datalog [25, 26]. ; for example, RT0 roles have no parameters, and SPKI/SDSI also lacks conjunctive policy rules. Li *et al.* show that SPKI/SDSI naming reduces to RT0 role-based trust [26], and that RT0 reduces to Datalog [25].

Other trust logics are merely syntactic variants of Datalog [8, 24], or like classical authorization logics are intractable. For example, FLANC [18]—recently proposed as a logic for authorizing network control in Software-Defined Exchanges (SDX)—is based on NAL [33, 36], which is undecidable. FLANC adds integer (flowspace) ranges to NAL, but these are already present in Datalog [25], and we illustrate their use in SAFE for secure routing.

### 8.2 Certificate Infrastructure

Using certificates with a decentralized key-value store (e.g., SafeSets) can be scalable, reliable, and secure (e.g., see [13]). For our use, the store must extend conventional K/V stores and DHTs to add write authorization (§5): neither clients nor SafeSets servers are trusted. We discuss issues with use of K/V stores or DHTs in this setting.

**Alternatives.** Note first that our approach is compatible with other options for handling logic content. Sets could be store and authenticated in other ways, e.g., in secure web directories maintained for the owning principals, or unsigned in a trusted metadata service for use within a single domain. This would require a different implementation of links, but certificates and even signing are no longer required. Trust logic and SAFE's scripted linking assume only that principals are authenticated to unique names: our use of public keys with hashing and signing meets this need, but they are not essential. Some settings may reduce cost by avoiding asymmetric crypto and authenticating principals by other means (e.g., Web single-sign on). Because slang abstracts set post/fetch, the certificate infrastructure is a replaceable element of SAFE.

**Revocation, renewal, and key rotation.** In any case, linking with on-demand fetch can assist with perennial problems for credential management. For example, an issuer may modify a certificate in SafeSets at any time, e.g., to renew it or revoke it. The caches use TTL consistency, so an authorizer may continue to use stale certificates until they expire. The expiration time is under control of the issuer, and an authorizer may pull/poll for a fresh copy at any time, as SAFE does when a logic query fails. A renewed certificate is automatically refreshed when it expires. Key rotation is more difficult because reissuing a certificate under a new key changes its token/link: any certificate that links to it must also be reissued (one hop).

**Integrity.** If a SafeSets server is subverted it can only deny service or leak data: it cannot forge certificates. A faulty server may hide updates or allow content to be overwritten: servers may be held strongly accountable (in the sense of CATS [37]) for such misbehavior.

**Confidentiality.** A faulty server may monitor access patterns or leak certificates without detection: we leave these issues to future work. A curious client may read any content for which it can synthesize the token (§3. It is easy to protect content by salting the label—which makes the token "unguessable"—but this also limits the ability of a peer to query for the certificate with a synthesized token, unless the issuer passes the salt to trusted peers out of band. Leaked tokens do not compromise integrity: it is not the token itself that confers privilege, but the authenticated statements contained in the logic set that it references.

**Malicious content.** Issuers may issue malformed certificates or generate malformed DAGs, e.g., by creating cycles in the DAG. The fetch procedure rejects malformed certificates and detects and ignores cycles. Valid certificates contain only logic statements, which share the safety properties of pure Datalog: all queries terminate and are sound with respect to trust policy rules, which identify and ignore any untrusted statements. However, issuers may create very large or costly logic sets to mount a denial of service attack. To defend against such attacks, authorizer may bound the size of incoming logic sets and query contexts at its discretion, and/or throttle its fetches.

**Reclamation.** Stored logic content may accumulate over time. SafeSets may delete any certificate after it expires. Even so, issuers may use unreasonable expiration times or simply post useless data to the store. SafeSets authenticates each issuer by its public key, but quotas are of no help if an issuer can mint new keys at will. One option is to apply a more stringent access check for posting (e.g., using SAFE).

## 9. Conclusion

Certificate linking resolves the credential discovery problem to enable practical use of trust logic in complex application scenarios. Scripted linking allows developers to generate small, efficient proof contexts with linear inference cost, given suitable indexing. Links enable fetching and caching of certificates on demand from a K/V store, which is inherently scalable. The primary metric for scalability then is the throughput at which any given participant can process logic queries, e.g., for access control decisions, given the complexity of the application's trust policies and certificate structures. The results show that linking enables efficient contexts via delegation-driven context pruning, enabling a logic server to issue queries against selected context sets at high throughput.

**Acknowledgements.** Vamsidhar Thummala implemented the initial version of the SAFE prototype, including the slang scripting language. Hang Zhu implemented the initial version of the **Attestation** scripts.